\documentclass[preprintnumbers,amsmath,amssymb,aps,preprint]{revtex4}

\usepackage{graphicx}
\usepackage{multirow}
\usepackage{tabularx}

\preprint{DESY 06-157}

\newcommand{\bee}{\begin{equation}}
\newcommand{\ee}{\end{equation}}
\newcommand{\beea}{\begin{eqnarray}}
\newcommand{\eea}{\end{eqnarray}}
\newcommand{\gfive}{\gamma_5}

\def\Tr{{\rm Tr}}

\begin{document}

\title{Quark condensate in two-flavor QCD}

\author{Thomas DeGrand,  Zhaofeng Liu}
\affiliation{
Department of Physics, University of Colorado,
Boulder, CO 80309 USA
}

\author{Stefan Schaefer}
\affiliation{
NIC, DESY Zeuthen,
Platanenallee 6,
D-15738 Zeuthen,
Germany
}

\begin{abstract}
We compute the condensate in QCD with two flavors of dynamical fermions using
numerical simulation. The simulations use overlap fermions, and the condensate
is extracted by fitting the distribution of low lying eigenvalues of the Dirac operator
in sectors of fixed topological charge to the predictions of Random Matrix Theory.
\end{abstract}

\maketitle

%%%%%%%%%%%%%%%%%%%%%%%%%%%%%%%%%%%%%%%%%%%%%%%%%%%%%%%%%%%%%%%%%%%%%%
\section{Introduction}
%%%%%%%%%%%%%%%%%%%%%%%%%%%%%%%%%%%%%%%%%%%%%%%%%%%%%%%%%%%%%%%%%%%%%%

The quark condensate $ \Sigma=\langle 0| \bar q q|0\rangle$ is the order parameter
associated with the spontaneous breaking of chiral symmetry in QCD. Along with the
pseudoscalar decay constant, it is 
one of the two fundamental parameters of the lowest order chiral Lagrangian which is the
low energy effective field theory for QCD. As such, its value is an interesting
physical quantity whose determination presents a challenge to lattice QCD methodology.
While there have been many calculations of $\Sigma$ in quenched QCD, results
for QCD with  dynamical fermions are relatively sparse
(for a summary of recent results, see Ref. \cite{McNeile:2005pd}).
This paper is a computation of $\Sigma$ for QCD with two flavors of dynamical overlap 
fermions\cite{Neuberger:1997fp,Neuberger:1998my}.
Overlap fermions implement chiral symmetry exactly at nonzero lattice 
spacing via the Ginsparg-Wilson\cite{Ginsparg:1981bj} relation.
This paper is an extension of our earlier work published in Ref.~\cite{DeGrand:2005vb}, 
which was primarily about algorithms, and Ref.~\cite{DeGrand:2006uy}, where
we use the same methodology to extract $\Sigma$ in $N_f=1$ QCD. 

Rather than measure $\langle \bar q q \rangle$ directly, we will determine
the  particular combination of the coefficients of the low energy effective field theory,
 $\Sigma= f^2 B$, in the usual parameterization
\bee
{\cal L}_2 =
\frac{f^2}{4}\Tr( \partial_\mu U \partial_\mu U^\dagger)+ B \frac{f^2}{2}\Tr[ M(U+U^\dagger)].
\ee
One expects that the quantity $\langle \overline q q\rangle$ (as computed, for 
example, in a lattice simulation at some quark mass $m_q$ and simulation volume $V$) is a 
function of $\Sigma$, $f$, $m_q$, and $V$. For the remainder of this paper, we 
will refer to the quantity $\Sigma$ as the condensate, and this is
the quantity we will compute on the lattice.

This is done using the  low-lying eigenvalues of the QCD Dirac operator in a finite
volume, whose distribution can be predicted by random matrix
theory (RMT)~\cite{Shuryak:1992pi,Verbaarschot:1993pm,Verbaarschot:1994qf}.
This hypothesis has been checked extensively by  lattice calculations,
mainly in quenched simulations
\cite{Berbenni-Bitsch:1997tx,Damgaard:1998ie,Gockeler:1998jj,Edwards:1999ra,Giusti:2003gf,Follana:2005km,Wennekers:2005wa},
but also in dynamical ones using staggered quarks
\cite{Berbenni-Bitsch:1998sy,Damgaard:2000qt}.
Our analysis is based on the distribution of the $k$-th eigenvalue from RMT as presented
in Refs.~\cite{Damgaard:2000ah,Damgaard:2000qt}. 
The prediction is for the distribution of the $k$-th  eigenvalue of the Dirac operator,
$\lambda_k$, in each topological sector, which is a function of the
dimensionless quantity $\zeta=\lambda_k \Sigma V$, where $\Sigma$ is the chiral 
condensate and $V$ is the volume of the box. These distributions are universal and 
depend only on the number of flavors and the topological charge. They depend
parametrically on  the dimensionless quantity $m_q \Sigma V$.
By comparing the distribution of the eigenvalues with the RMT
prediction one can thus measure the chiral condensate $\Sigma$.
This method gives the zero quark mass, infinite volume condensate directly.

The validity of the approach can be verified by comparing the shape of the 
distribution for the various modes and topological sectors. A too small 
volume  causes deviations in the shape, particularly for the higher modes.
Two recent large scale studies using the overlap operator on quenched
configurations~\cite{Bietenholz:2003mi,Giusti:2003gf},  found that 
with a length larger than $1.2~{\rm fm}$ and $1.5~{\rm fm}$ respectively,
the RMT predictions match the result of the
simulation. Our dynamical lattices have a spatial extent of about 1.5~fm
and we have a smaller volume from our earlier work.
As we will see, random matrix theory describes our data quite well.

The RMT predictions are made with the assumption that the volume is infinite.
In the epsilon-regime
of chiral perturbation theory, finite volume modifies the formula for the condensate
by multiplication by a shape factor, $a^3\Sigma \rightarrow \rho a^3\Sigma$,
where
\bee
\rho = 1 + \frac{N_f^2-1}{N_f}\frac{c(l_i/l)}{f_\pi^2L^2}
\label{eq:rho}
\ee
and $c(l_i/l)$ depends on the geometry\cite{Gasser:1986vb}. (It is 0.1405 for
hypercubes.) We do not know $\rho$ since we have
not measured $f_\pi$, but combining our lattice spacing and lattice size with $f_\pi=93$~MeV,
we expect $\rho \sim 1.42$ for our simulations.
In what follows, we will refer to the quantity we extract from RMT fits as $\Sigma$
rather than the more proper label of $\rho \Sigma$. Ignoring $\rho$ introduces
a systematic uncertainty into our result.

The lattice calculation has four parts: The first is the algorithm.
Our calculations are performed with overlap fermions, which possess exact chiral symmetry
at nonzero lattice spacing.  A variation on the standard Hybrid Monte Carlo allows
us to perform simulations in sectors of fixed topology\cite{Bode:1999dd,Cundy:2005mr,ouralg}. 
We review the algorithm in Sec.~\ref{sec:lattaction}.

The extraction of $\Sigma$ from a fit to the eigenvalues of the Dirac operator 
is described in Sec. \ref{sec:fitting}.

Next, we need a lattice spacing to convert the dimensionless lattice-regulated condensate
to a dimensionful number. We obtain the lattice spacing through the Sommer parameter
from the static quark potential\cite{Sommer:1993ce}.
This is described in Sec. \ref{sec:vr}.

Finally, we need a matching factor, to convert the lattice-regulated condensate to its
$\overline{MS}$ value. We do this using 
the Regularization Independent scheme\cite{Martinelli:1994ty}.
This is described in Sec. \ref{sec:ri}.

To anticipate our results, which are summarized in Sec. \ref{sec:results}, the
larger simulation volume considerably improves the quality of the RMT fits. The value
of the lattice-regulated condensate we obtain here is quite consistent with
 our earlier result. We find, however, that the nonperturbative matching factor we
determine is quite different from the perturbative matching factor used in 
Ref. \cite{DeGrand:2005vb}.
Our final answer is $\Sigma^{1/3} = 282(10)$ MeV; the quoted error is statistical 
(from the simulation);
in addition, we estimate that this number exceeds the true result by a systematic
factor $\rho^{1/3}=1.13$.

%%%%%%%%%%%%%%%%%%%%%%%%%%%%%%%%%%%%%%%%%%%%%%%%%%%%%%%%%%%%%%%%%%%%%%
\section{Methodology\label{sec:method}}
%%%%%%%%%%%%%%%%%%%%%%%%%%%%%%%%%%%%%%%%%%%%%%%%%%%%%%%%%%%%%%%%%%%%%%

\subsection{Lattice action and simulation algorithm \label{sec:lattaction}}

The massless overlap\cite{Neuberger:1997fp,Neuberger:1998my} operator is
\bee
D =D_{ov}(m=0)= R_0 \left[ 1+\gfive \epsilon(h(-R_0))\right]
\label{eq:Dov}
\ee
where $\epsilon(h)=h/\sqrt{h^2}$ is the sign function of the Hermitian kernel
operator $h=\gfive d$ which is taken at  negative mass $R_0$.
The squared Hermitian overlap operator
$H^2=(\gamma_5 D)^2=D^\dagger D$
commutes with $\gamma_5$ and therefore can have eigenvectors with
definite chirality. The modes at zero and $4R_0^2$ aside (which are associated with
zero modes of $D$), the spectrum is doubled
with a positive and a negative chirality eigenvector for eigenvalue
$|\lambda|^2$.

The RMT analysis requires data restricted to particular topological sectors.
It is very convenient to generate these data sets directly, rather than
letting the topology vary in the simulation and filtering it into different
topological
sectors. A simple observation\cite{Bode:1999dd,Cundy:2005mr,ouralg} allows us to do this: the
fermion determinant for each  flavor is equal to the determinant of $H^2$
in one chiral sector times a correction factor for the modes at $m$ and $2R_0$:
\bee
{\det} D = (m/2R_0)^{|Q|} {\det} H_{\rm opp}^2
\label{eq:replace}
\ee
where $H_{\rm opp}^2$ is the squared Dirac operator in the chiral sector without zero modes.
Since  $H_{\rm opp}^2$ is a positive operator, its determinant  can be
replaced by a  pseudofermion estimator\cite{ouralg} involving a single chiral pseudofermion
for each quark flavor.
Then an ensemble can be generated using Hybrid Monte Carlo\cite{Duane:1987de}.

At topological boundaries, the spectrum of the Dirac operator is discontinuous,
and so is the fermionic contribution to the action. In the algorithm of
  Ref.~\cite{Fodor:2003bh}, the molecular dynamics trajectory either ``reflects''
from the boundary, with no topological change, or ``refracts'' and changes its
topology. We are going to use the algorithm for simulations
on sectors of fixed topology, so we simply forbid refractions. We pick chiral sources either in the
opposite chirality (if $Q\ne 0$) or randomly select a source chirality (if $Q=0$).
For other parts of the calculation, where we need data sets which are not restricted
to particular topological sectors, we use our implementation of the algorithm of
  Ref.~\cite{Fodor:2003bh}.

We now mention specific features of the simulation.
Our particular implementation of the Hybrid Monte Carlo algorithm has been
previously discussed in Refs.~\cite{DeGrand:2004nq,DeGrand:2005vb,Schaefer:2005qg}.
 We use the L\"uscher--Weisz gauge
action~\cite{Luscher:1984xn} with the  tadpole improved coefficients of
Ref.~\cite{Alford:1995hw}. Instead of determining the fourth root
of the plaquette expectation value $u_0=(\langle U_{pl}\rangle/3)^{1/4}$
self-consistently, we set it to 0.86 for all our runs as we did in our previous
publications.
 We are using a planar kernel Dirac operator $d$ with nearest and
 next-to-nearest (``$\sqrt{2}$") interactions. The choice of coefficients,
clover term, and value of $R_0$ are those
 of Refs.~\cite{DeGrand:2004nq,DeGrand:2005vb,Schaefer:2005qg}.
Our kernel operator $d$ is constructed from gauge links to
which two levels of isotropic stout blocking~\cite{Morningstar:2003gk} have
been applied. The  blocking parameter $\rho$ is set to 0.15.
 The sign function is computed using the
Zolotarev approximation
with an exact treatment of the low-lying eigenmodes $|\lambda\rangle$ of $h(-R_0)$.
We removed 16 kernel eigenmodes, which typically extended in magnitude
up to about 0.25 and allowed us to restrict the range of the
Zolotarev approximation from 0.9 of the maximum value to 2.7 (the upper limit for
eigenvalues of
our kernel action).

We simulate on $10^4$ lattices at one value of the gauge coupling $\beta=7.2$
(which we chose to be roughly at the $N_t=6$ phase transition), with
three values of the bare sea quark mass $am_q= 0.015$, $0.03$ and
$0.05$.

At each mass value and for $|Q|=0,1$ we ran several independent data streams:
three $Q=0$ streams for each mass, three $|Q|=1$ streams at $am_q=0.05$, and two
 $|Q|=1$ streams at $am_q=0.03$ and 0.015. Each stream ran for about 150 trajectories.
We used one extra level of Hasenbusch preconditioning\cite{Hasenbusch:2001ne}
 and typically broke
the time intervals into eight steps with the lightest mass, four to six steps with the
heavier mass, and 12 steps of Sexton-Weingarten integration for the gauge fields.
Our acceptance rated averaged about 70  per cent.
The whole simulation consumed about 120 processor-months on our array of 3 Ghz
P4E processors. We dropped the first 50 trajectories from each stream for thermalization
(and checked that our results were independent of this selection).
We recorded eigenvalues after every fifth HMC trajectory.

By running in the opposite chirality sector, we nearly eliminated critical slowing down:
the elapsed time for a trajectory was nearly independent of $Q$ and was only marginally
slower at $am_q=0.015$ than at $am_q=0.05$. This would not have been the case had
we used a conventional (non-chiral) HMC algorithm, because of the
near-zero mode.

This calculation was about a factor 2.5 times slower than our $N_f=1$ simulation
described in Ref. \cite{DeGrand:2006uy},
in terms of the cost of the application of an overlap operator
to a trial vector. (We did that calculation after we collected the data for this
project.) This is due to the lower level of stout smearing
(two steps with $\rho=0.15$ versus three steps for $N_f=1$.)

\subsection{Data fitting \label{sec:fitting}}
We computed the lowest four eigenvalues $|\lambda|^2$ of the squared Dirac operator $H^2$. 
From those we get the eigenvalues of the overlap operator $D$. These  lie on a circle,
which we project onto the imaginary axis via a M{\"o}bius transform
\bee
\tilde\lambda=\frac{\lambda}{1-\lambda/(2R_0)}.
\ee
The distribution of these eigenvalues is predicted by RMT in terms of one parameter $\Sigma$. 
Because we are at finite volume, however, this prediction is valid only for the lowest 
modes. From which level on the deviation between the observed data and the RMT curves
becomes significant is a priori not clear. Basically, RMT makes two predictions: The
positions of the peaks are equally spaced and it also predicts the shape of the distribution
of each individual level. 

These two predictions should be met by the data in order for the extraction
of $\Sigma$ to be meaningful.
The quality of the fit can be measured by the confidence level given by the 
Kolmogorov-Smirnov test \cite{NR}. It compares the integrated 
distributions (the cumulants) of the measured data $C(x)$ and the 
theoretical prediction $P(x)$.
The cumulant of the measured data is
$C(x) = n(x)/N$ where $n(x)$ is the number of data points with a value smaller 
than $x$ and $N$ the total number of data points.
The theoretical prediction for this quantity can be computed by integrating
the distribution: $P(x)=\int_{-\infty}^x f(y) {\rm d} y$.
This eliminates the bias introduced by binning the data and comparing it to the 
theoretical distribution.
The quantity of interest is the largest deviation of $P$ and $C$: 
$D=\max_x |P(x)-C(x)|$. From this the confidence level is given by
\bee
Q_{KS}\left((\sqrt{N}+0.12+0.11/\sqrt{N})D\right)
\ee
with the function $Q_{KS}$ given by
\bee
Q_{KS}(x)= 2\sum_{j=1}^\infty (-)^{j-1}\exp(-2j^2 x^2)\, .
\ee

\begin{figure}
\includegraphics[width=0.6\textwidth,angle=-90]{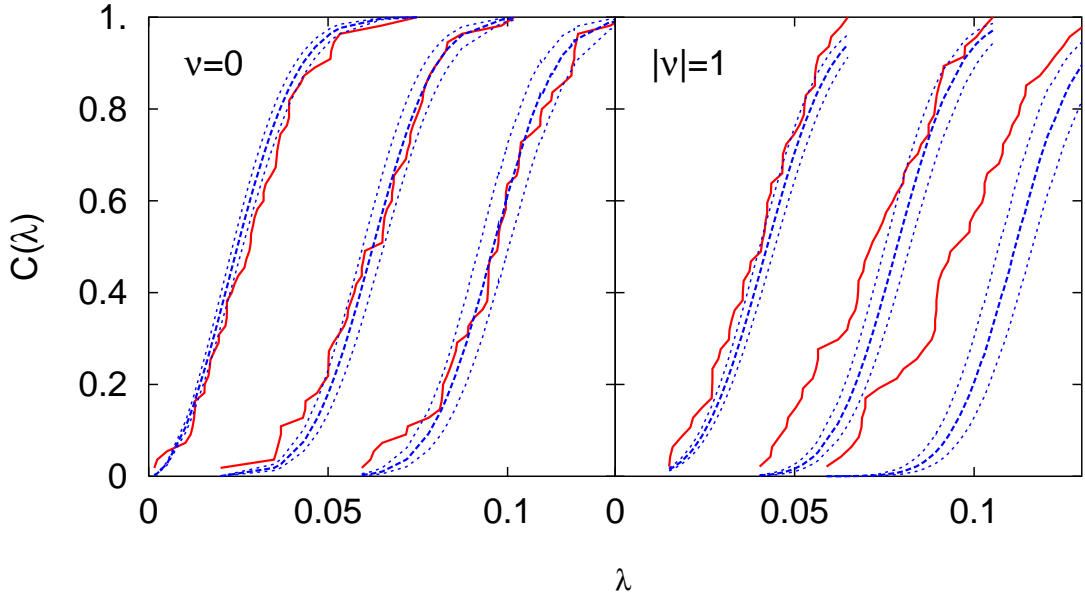}
\caption{The measured cumulants for the lowest three modes in $\nu=0$ (left) 
and $|\nu|=1$ (right).
The smooth dashed curve represents the RMT prediction with the value of $\Sigma$ from
the combined fit to the lowest mode in each sector. The dotted curves indicate the
$1\sigma$ range of $\Sigma$ determined by the bootstrap method.
\label{fig:cumul}
}
\end{figure}

Let us now start by discussing the $am_q=0.05$ data because of its larger statistics.
In Fig.~\ref{fig:cumul}, the solid line shows the measured cumulant for the 
lowest three modes in $\nu=0$ and $|\nu|=1$. This is to be compared to the 
integrated RMT prediction, which depends on the parameter $\Sigma$. We fit $\Sigma$
by maximizing the product over the confidence levels for each cumulant included
in the fit. It is a priori not clear for how many modes the RMT description is
valid. Due to the finite volume, the distributions are expected to deviate for the 
higher levels. We therefore try several combinations of the number of modes included in
the fits which we now discuss. The results of these fits can be found in Table~\ref{tab:CL10}.

\begin{table}
\begin{tabular}{c c c c|c cc}
  \hline
 $m$ & $a^3\Sigma$ &  $|\nu|$ & $N$   & CL level 1 & CL level 2 & CL level 3 \\
 \hline
  \hline
  \multirow{2}{*}{0.015 \ \ }& \multirow{2}{*}{0.0130(7)}& 0 &\ \ 26\ \ & \fbox{0.82} & 0.86 & 0.02  \\
                             &                           & 1 &\ \ 32\ \ & \fbox{0.42} & 0.15 & 0.71  \\
 \hline
  \multirow{2}{*}{0.03 \ \ }& \multirow{2}{*}{0.0112(7)} & 0 &\ \ 33\ \ & \fbox{0.12} & 0.57 & 0.04  \\
                             &                           & 1 &\ \ 28\ \ & \fbox{0.20} & 0.13 & 0.02  \\
 \hline
  \multirow{2}{*}{0.05 \ \ }& \multirow{2}{*}{0.0105(4)} & 0 &\ \ 55\ \ & \fbox{0.55} & 0.77 & 0.94  \\
                             &                           & 1 &\ \ 47\ \ & \fbox{0.64} & 0.01 & 0.00  \\
 \hline
 \hline
  \multirow{2}{*}{0.05 \ \ }& \multirow{2}{*}{0.0098(5)} & 0 &\ \ 55\ \ & \fbox{0.52} & 0.79 & 0.93  \\
                             &                           & 1 &\ \ 47\ \ & 0.67 & 0.02 &       \\
 \hline
 \hline
  \multirow{2}{*}{0.05 \ \ }& \multirow{2}{*}{0.0111(6)}& 0 &\ \ 55\ \ & 0.07 & 0.03 & 0.001  \\
                            &                           & 1 &\ \ 47\ \ & \fbox{0.99} & 0.08 & 0.0005      \\
\hline
 \hline
  \multirow{2}{*}{0.05 \ \ }& \multirow{2}{*}{0.0106(4)} & 0 &\ \ 55\ \ & \fbox{0.52} & \fbox{0.80} & 0.90  \\
                             &                           & 1 &\ \ 47\ \ & \fbox{0.67} & 0.02 & 0.00   \\
 \end{tabular}
 \caption{$a^3\Sigma $ from a fit to the lowest eigenvalues on the $10^4$ ensembles. The confidence levels for the 
 individual distributions are given in the last three columns; they are boxed if the level is included
 in the fit.
  \label{tab:CL10}}
\end{table}

\begin{figure}

\includegraphics[width=0.45\textwidth,clip,angle=-90]{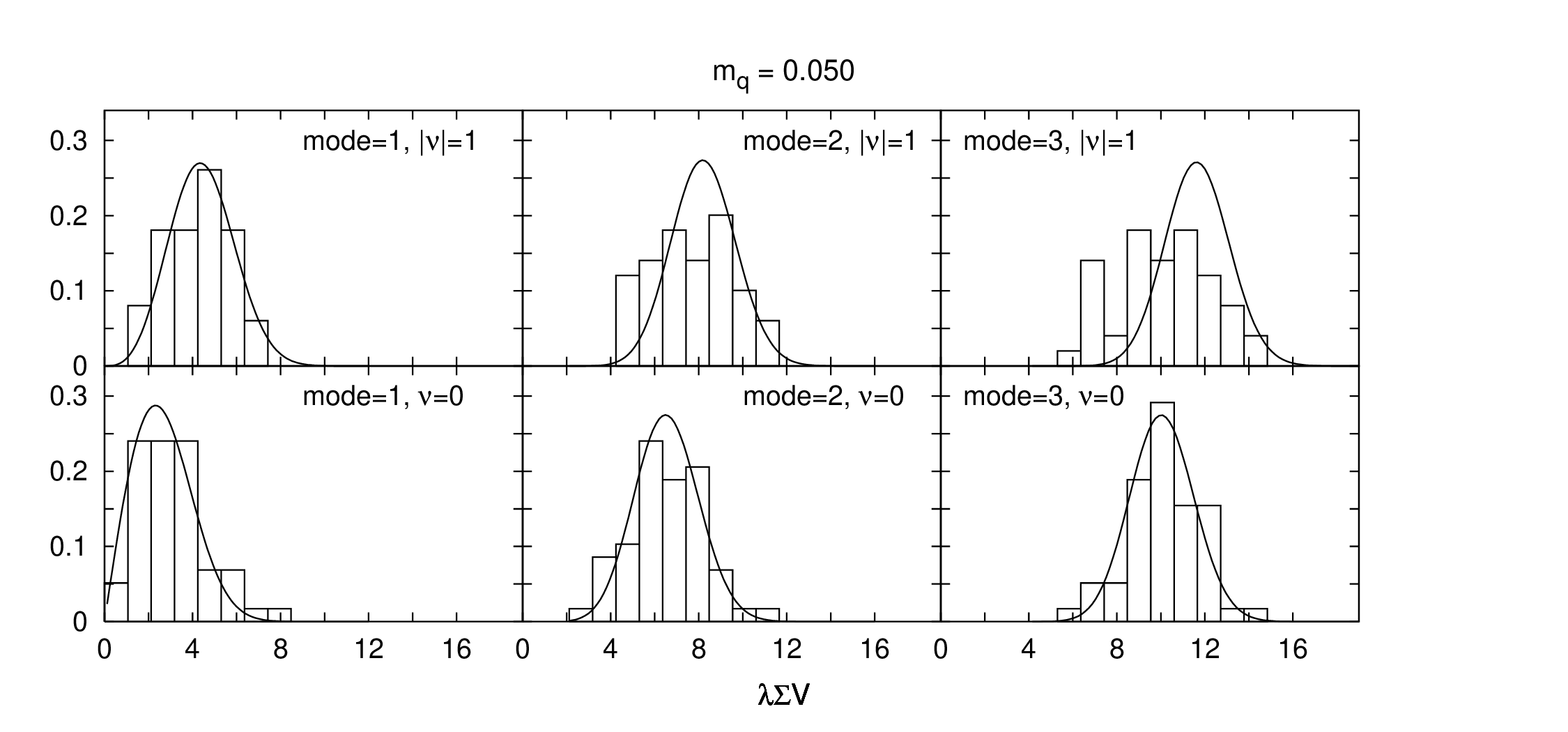}

\includegraphics[width=0.45\textwidth,clip,angle=-90]{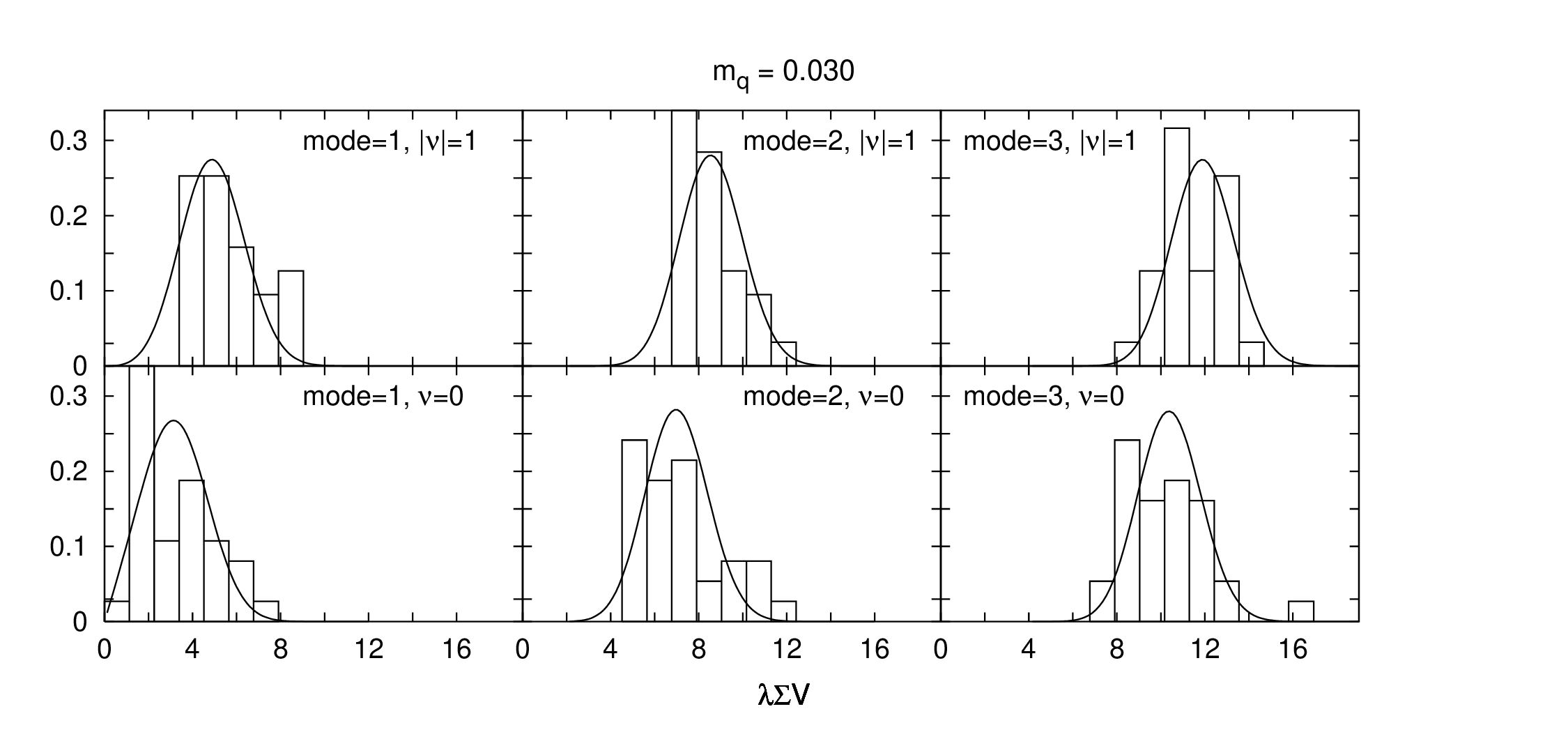}

\includegraphics[width=0.45\textwidth,clip,angle=-90]{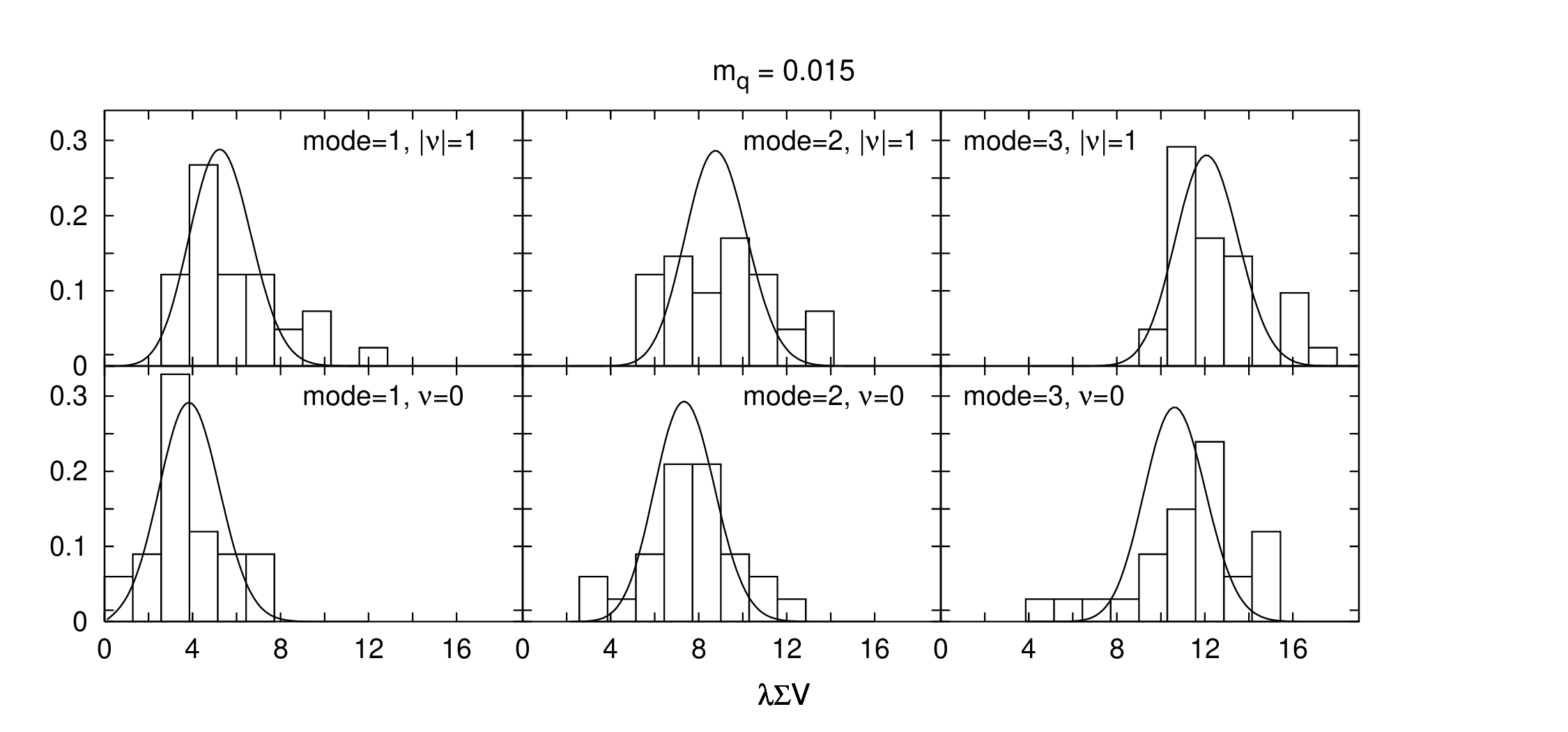}

\caption{RMT fit to eigenvalue distributions for $am_q=0.05$,  $0.03$ and $0.015$. The value
of $a^3\Sigma$ is determined from a combined fit to the lowest mode in each 
topological sector.\label{fig:dist}}
\end{figure}

The result of a combined fit to the lowest mode in each
topological sector is shown in Fig.~\ref{fig:cumul}. 
To visualize the range of the uncertainty in $a^3\Sigma=0.0105(4)$, we 
show the theoretical curves for the two extremal values of the one sigma
range. The errors on the fit parameter are determined by the 
bootstrap procedure.
(See Fig.~\ref{fig:dist} (top) for a comparison of the binned data with the  RMT distribution.)
We find very good agreement for the shape and position of the lowest three modes in $\nu=0$.
In $\nu=1$ however, only the lowest mode matches. For all these modes, the confidence levels
are above $50\%$. See Table~\ref{tab:CL10} for the confidence levels of the
fits to the individual distributions.

To check for consistency between the two topological sectors, 
we show the results from separate fits of the RMT prediction to the
distribution in each topological sector. 
The general statements from the combined fit  persist: 
The lowest three modes in $\nu=0$ and the lowest one in 
$|\nu|=1$ can be described by RMT. The extracted values of $a^3\Sigma$ agree within
uncertainties (the  $\nu=0$ ensemble is completely independent from the $|\nu|=1$ ensemble).
We get $a^3\Sigma=0.0098(5)$ from $\nu=0$ and  $a^3\Sigma=0.0111(6)$ from $\nu=1$.

So far we have focused on the $am_q=0.05$ data set, because of its higher statistics.
The streams at  $am_q=0.03$ and 0.015, however, give us the possibility to 
check, whether the quark mass dependence of the RMT prediction is correct.
From a combined fit to the distribution of the lowest eigenvalue
in each topological sector 
(the fit we also chose to use for our final value for $am_q=0.05$),
we extract $a^3\Sigma=0.0112(7)$ and $a^3\Sigma=0.0130(7)$
for  $am_q=0.03$ and 0.015 respectively.
Fig.~\ref{fig:dist} shows the RMT predictions from these values of $\Sigma$ along with the
measured distributions.
   
\begin{figure}
\includegraphics[width=0.5\textwidth]{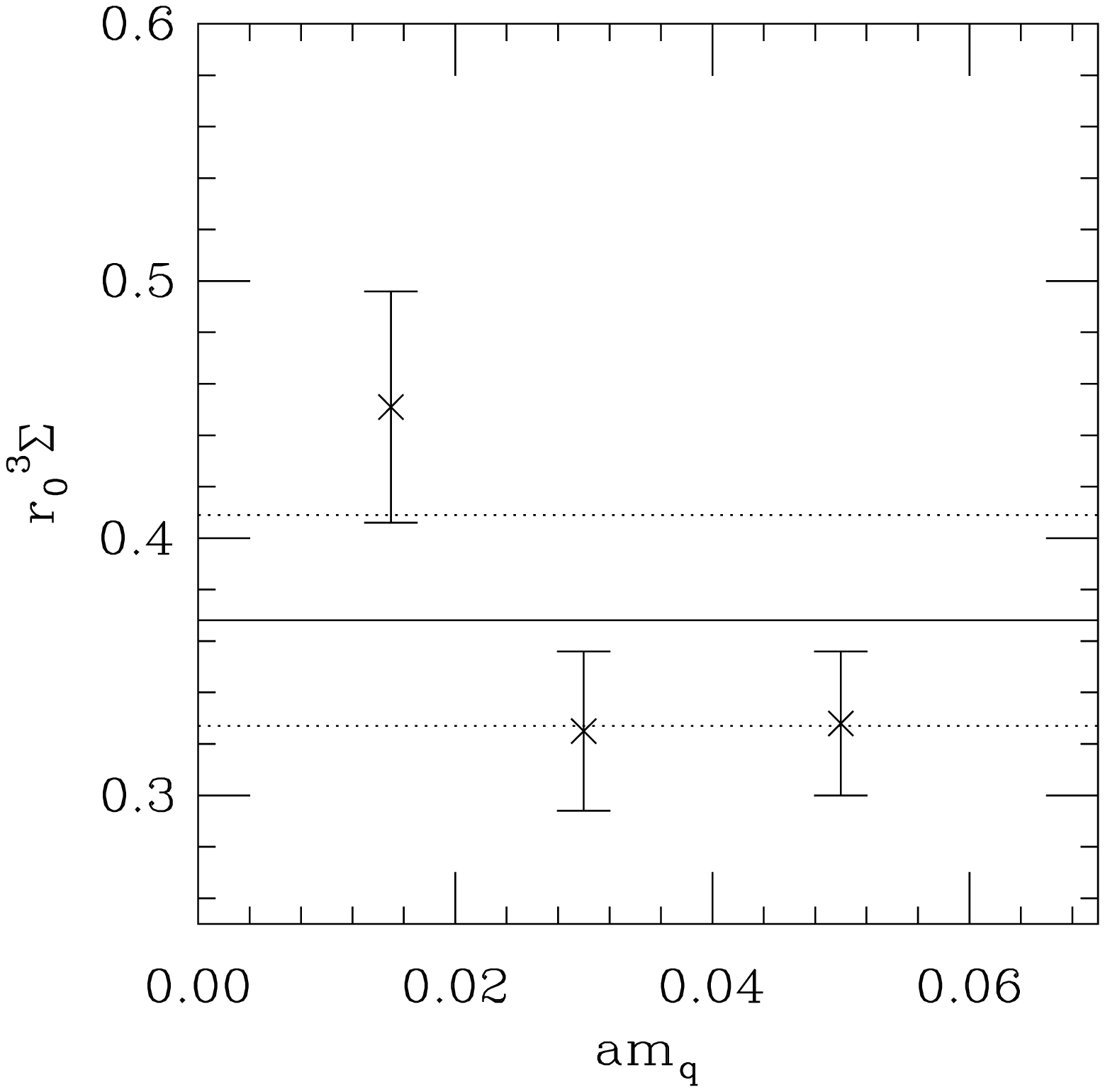}
\caption{\label{fig:r0sigma} The dimensionless quantity $r_0^3\Sigma$ in $\overline{MS}$ scheme at $\mu=2$ GeV,
as a function
of the bare quark mass. The horizontal lines represent the averaged value $r_0^3\Sigma=0.368(41)$
and its uncertainty.}
\end{figure}

The values of $a^3\Sigma$ for the three
sea quark masses seem to be only barely compatible, however, one has to compare them
with the difference in the lattice spacing taken into account. Using the results from 
Sec.~\ref{sec:vr} and Sec.~\ref{sec:ri} (to convert our numbers to
$\overline{MS}$ regularization scheme) we plot $r_0^3\Sigma$ as a function of the bare sea quark mass and
obtain a reasonable  agreement within uncertainties, see Fig.~\ref{fig:r0sigma}.
Taking the average of the three values we find $r_0^3\Sigma=0.368(41)$.

The measured confidence levels are much better than what we found on smaller lattices
which makes us confident that the deviations which we observe are only due to
the finite volume.
In Ref. \cite{DeGrand:2005vb}, we simulated on $8^4$ lattices with a same lattice
spacing. The volume there was clearly too small.  To demonstrate the 
improvement due to the larger volume, we have re-analyzed the data with
the techniques described above. The results are given in Table~\ref{tab:CL8} and
should be compared to the second and third entry in  Table~\ref{tab:CL10}, because in both cases  
we fitted to the lowest mode in each topological sector.
We clearly observe an improvement in the confidence level when we go to 
the larger volume, i.e.  the RMT prediction match the measured cumulants much better.

We recall (see Eq.~\ref{eq:rho})
that the result $\Sigma$ of finite simulation volume is equal to $\rho
\Sigma_\infty$, where $\Sigma_\infty$  is the infinite volume condensate.
 We do not know $\rho$ since we have
not measured $f_\pi$, but combining our lattice spacing and lattice size with $f_\pi=93$~MeV,
we expect $\rho \sim 1.42$ for $L=10$ and $\rho(L=10)/\rho(L=8)=0.86$. 
In order to compare the results from the two volumes, we have to divide each of them by its
respective
$\rho$. For $m=0.05$ we therefore
obtain for the $10^4$ lattices $a^3  \Sigma_\infty=0.0074(3)$ and for the $8^4$ lattices
$a^3  \Sigma_\infty=0.0080(3)$. The error-bars of the two values touch. However, we suspect that
the $8^4$ data
may be too small for RMT to provide a reliable determination of the condensate.
We are unwilling to attempt a combined fit of $\rho \Sigma$ in terms of $f_\pi$ and $\Sigma$ from data obtained on several 
simulation volumes without good RMT fits for each volume.

\begin{table}
\begin{tabular}{c c c c|c cc}
  \hline
 $m$ & $a^3\Sigma$ &  $|\nu|$ & $N$   & CL level 1 & CL level 2 & CL level 3 \\
 \hline
 \hline
  \multirow{2}{*}{0.03 \ \ }& \multirow{2}{*}{0.0124(5)} & 0 &\ \ 112\ \ & \fbox{$6\cdot10^{-5}$} & 0.06 & 0.05  \\
                            &                            & 1 &\ \ 58\ \ & \fbox{$9\cdot10^{-5}$} &  $0.002$ & $1\cdot10^{-5}$   \\
 \hline
  \multirow{2}{*}{0.05 \ \ }& \multirow{2}{*}{0.0132(5)}  & 0 &\ \ 96\ \ & \fbox{0.01} & 0.004 & $0.03$  \\
                            &                             & 1 &\ \ 75\ \ & \fbox{0.04} & 0.003 & 0       \\
 \end{tabular}
 \caption{To demonstrate the effect of the finite volume, we also re-analyzed our old $8^4$ data. 
 The notation is the same as in Tab.~\ref{tab:CL10}.
 This table corresponds to the second and third entry in  Tab.~\ref{tab:CL10} for the $10^4$ lattices. 
 The confidence levels are considerably lower.
 \label{tab:CL8}}
\end{table}

\subsection{Lattice spacing \label{sec:vr}}
We determined an overall scale from a fit to the static quark potential. It is
extracted from the effective masses
of Wilson loops after one level of HYP smearing~\cite{Hasenfratz:2001hp,Hasenfratz:2001tw},
where the short-distance effects of
the HYP smearing are corrected using a fit to the perturbative lattice
artifacts. We measured the potential on 76, 82, and 48 $8^3\times12$ configurations
at $am_q=0.05$, 0.03, and 0.015 (spaced 5 HMC trajectories apart, from several streams per mass).
 To generate these data sets we used a 
conventional two-flavor HMC algorithm.
Fits with the minimum distance $t=4$ and 5 are consistent within uncertainties. 
In Ref.~\cite{DeGrand:2005vb} we only had $8^4$ lattices, which prevented us from going to 
these separations. The determinations of $r_0/a$ from those data sets are slightly different
from what we have here, and were slightly contaminated by excited states.
A compilation of fit results is shown in Table \ref{tab:R0}.

\begin{table}
\begin{tabular}{c|c|c}
\hline
$m_q$      &     $r_0/a$     &      $a$[fm]       \\
\hline
0.015  \ \  & \ 3.47(8)  \  & \ 0.144(3)     \\
0.03  \ \  & \ 3.27(6)  \  & \ 0.153(3)    \\
0.05  \ \  & \ 3.35(7) \   &   0.148(3)     \\
\end{tabular}
\caption{\label{tab:R0}The Sommer parameter $r_0$ and the lattice spacing $a$ in fm from $r_0=0.5~{\rm fm}$.
We show results from two fit ranges used to extract the potential. The fit range for the fit to the
potential was $r\in[1.4,6.1]$ with little variation between different choices for
this range. }
\end{table}

%%%%%%%%%%%%%%%%%%%%%%%%%%%%%%%%%%%%%%%%%%%%%%%%%%%%%%%
\subsection{Conversion to $\overline{MS}$ regularization \label{sec:ri}}
A renormalization constant is needed to convert the lattice
result of the quark condensate to its $\overline{MS}$ value.
To get this matching factor, we use the RI' scheme introduced in
Ref.~\cite{Martinelli:1994ty}. Specifically we follow the procedure 
described in Ref.~\cite{DeGrand:2005af}. The RI' scheme result
in the chiral limit can be
converted to the 2 GeV $\overline{MS}$ value by using the ratio
connecting the two schemes. The ratio was computed by continuum
perturbation theory to three loops~\cite{Franco:1998bm, Chetyrkin:1999pq}.

The simulation for getting the matching factor should not be restricted 
in topological sectors.
%To produce a matching factor, 
Also, one needs simulations with a momentum scale
short enough to be free from nonperturbative effects and yet not too 
short to be affected
by discretization effects. 
Somewhat to our surprise, our sample of $8^4$ lattices
proved suitable to do this. We
combined multiple streams of data together to reduce simulation-time
autocorrelations. Our data sets consisted of approximately 50 configurations
per mass value at two values of the bare sea quark mass, $am_q=$0.03 and
0.05.

The $8^4$ lattice is periodic in space directions and
antiperiodic in the time direction. Therefore the momentum values are
\begin{equation}
ap_\mu=\left(\frac{2\pi}{8}k_x,\frac{2\pi}{8}k_y,\frac{2\pi}{8}k_z,
\frac{\pi}{8}(2k_t+1)\right).
\end{equation}
We choose the values of $k_\mu$ such that
the momentum values lie
as close as possible to the diagonal of the Brillouin zone. The maximum
value of $ap=2.115$ corresponds to $k_\mu=(2,1,0,1)$. The 
propagators are cast from a point source and then  projected
to the desired momentum values.

\begin{figure}
\begin{center}
\includegraphics[width=0.6\textwidth,clip]{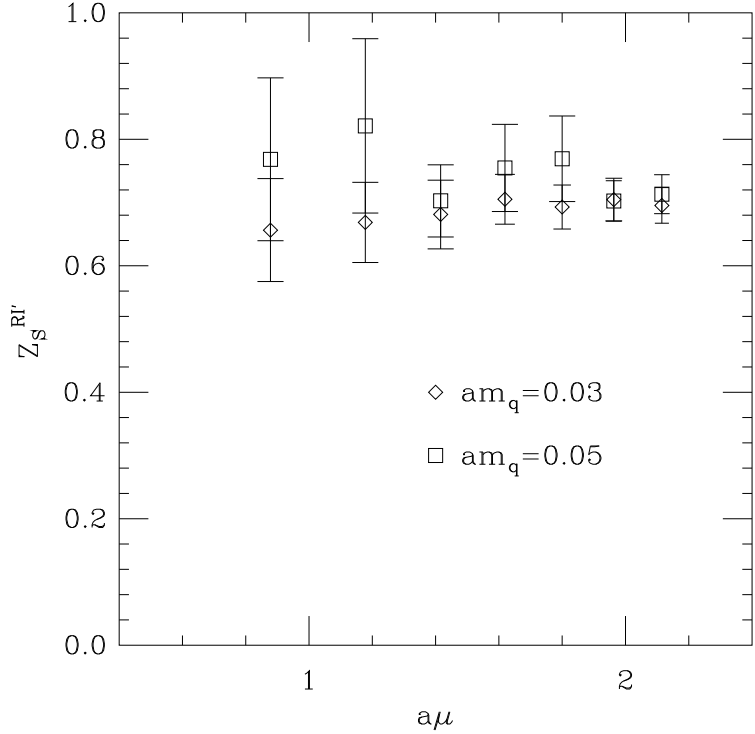}
\end{center}
\caption{$Z_S^{RI'}$ for quark mass $am_q=0.03$ and 0.05.}
\label{fig:zs}
\end{figure}

The renormalization constant for the scalar density in the RI' scheme 
is given in Fig.~\ref{fig:zs}.
Values of $Z_S^{RI'}$ for $am_q=0.03$ and 0.05 are listed in
Table~\ref{zstable}. The inverse lattice spacings are 1.29 GeV and 1.32 GeV
for the
$am_q=0.03$ and 0.05 data sets determined
%~\cite{DeGrand:2005vb}
from the Sommer parameter. Thus $\mu=2$~GeV corresponds to $a\mu=1.55$
or 1.52. The 2~GeV RI' values are obtained from
linear
interpolations from the two closest $\mu$ values of the data.
\begin{table}
\caption{Values of $Z_S$ in the RI' scheme for the two quark masses.
The inverse lattice spacings are 1.29 GeV and 1.32 GeV for the
$am_q=0.03$ and 0.05 data sets respectively from the
Sommer parameter. Therefore $\mu=2$~GeV corresponds to $a\mu=1.55$
or 1.52 accordingly. The 2 GeV RI' values are obtained from
linear
interpolations from the two closest $\mu$ values of the data.}
\begin{center}
\begin{tabular}{cll}
\hline\hline
$a\mu$ & $am_q=0.03$ & $am_q=0.05$ \\
\hline
0.878  &0.66(8)      &0.77(13)  \\
1.178  &0.67(6)      &0.82(14) \\
1.416  &0.68(5)      &0.70(6)  \\
1.619  &0.71(4)      &0.75(7)  \\
1.800  &0.69(3)      &0.77(7)  \\
1.963  &0.70(3)      &0.70(3)  \\
2.115  &0.70(3)      &0.71(3)  \\
$\mu=2$GeV & 0.70(4) & 0.73(6) \\
\hline\hline
\end{tabular}
\end{center}
\label{zstable}
\end{table}

The conversion ratio for the scalar and pseudoscalar densities, in
Landau gauge and to three loops, is~\cite{Franco:1998bm,
Chetyrkin:1999pq}
\begin{eqnarray}
\frac{Z_S^{\overline{\rm MS}}}{Z_S^{\rm RI'}}&=&
\frac{Z_P^{\overline{\rm MS}}}{Z_P^{\rm RI'}}
=1+\frac{16}{3}\frac{\alpha_s}{4\pi}+\left(\frac{4291}{18}
-\frac{83n_f}{9}-\frac{152\zeta_3}{3}\right)
\left(\frac{\alpha_s}{4\pi}\right)^2
\nonumber \\
&&+\left(\frac{3890527}{324}-\frac{241294n_f}{243}
+\frac{7514n_f^2}{729}-\frac{224993\zeta_3}{54}\right.\nonumber\\
&&+\left.\frac{4720\zeta_3n_f}{27}
+\frac{32\zeta_3n_f^2}{27}
-\frac{80\zeta_4n_f}{3}
+\frac{2960\zeta_5}{9}\right)\left(\frac{\alpha_s}{4\pi}\right)^3
+O(\alpha_s^4),
\label{zspratio}
\end{eqnarray}
where $n_f$ is the number of flavors and $\zeta_n$ is the Riemann zeta 
function evaluated at $n$. This formula is for zero quark mass.

To get numerical results of the above ratio, we
use the coupling constant from the so-called ``$\alpha_V$" scheme.
As in the appendix of Ref.~\cite{DeGrand:2005vb}, from the one-loop
expression relating the plaquette to the coupling
\begin{equation}
\ln\frac{1}{3}\Tr U_p=-\frac{8\pi}{3}\alpha_V(q^*)W,
\end{equation}
where $W=0.366$ and $q^*a=3.32$ for the tree-level L\"uscher--Weisz
action, we obtain $\alpha_V(3.32/a)=0.192$ and 0.193 for the
$am_q=0.03$ and 0.05 data sets. 
Then $\alpha_s^{\overline{MS}}(e^{-5/6}3.32/a)$ is
calculated~\cite{Brodsky:1982gc} and run to 
$\alpha_s^{\overline{MS}}$(2 GeV) by using
$\beta_0=29/12\pi$ and $\beta_1=230/48\pi^2$ for two flavor QCD. 
We find $\alpha_s^{\overline{MS}}$(2 GeV)$=0.210$ and 0.213 for the
$am_q=0.03$ and 0.05 data sets respectively. Then from
Eq.(\ref{zspratio}) with $n_f=2$, we get
$Z_S^{\overline{MS}}/Z_S^{RI'}=1.160$ and 1.164 accordingly.
Therefore $Z_S^{\overline{MS}}$(2 GeV)$=0.81(5)$ and 0.85(7) for the two
data sets. 
Because these two values have rather large uncertainty, we perform
the chiral limit by fitting to a constant and get $Z_S^{\overline{MS}}$(2~GeV)=0.83(4).

%%%%%%%%%%%%%%%%%%%%%%%%%%%%%%%%%%%%%%%%%%%%%%%%%%%%%%%%%%%%%%%%%%%%%%
\section{Results \label{sec:results}}
%%%%%%%%%%%%%%%%%%%%%%%%%%%%%%%%%%%%%%%%%%%%%%%%%%%%%%%%%%%%%%%%%%%%%%
Combining our results for the lattice regulated condensate, the Sommer parameter, and the
Z-factor, we find
$r_0^3 \Sigma(\overline{MS},\mu=2 \ {\rm GeV})=0.328(28)$, 0.325(31), and 0.451(45) at
$am_q=0.05$, 0.03, and 0.015 respectively. These values are plotted in Fig.~\ref{fig:r0sigma}.
 The result from each quark mass
is the parameter of the Chiral Lagrangian. Averaging them, we find
\bee
r_0^3 \Sigma (\overline{MS},\mu=2 \ {\rm GeV}) = 0.368(41).
\ee
Taking the real-world value for $r_0=0.5$ fm, this is
\bee
\Sigma(\overline{MS},\mu=2 \ {\rm GeV}) = 0.0225(25)  {\rm GeV}^3
\ee
or
\bee
( \Sigma(\overline{MS},\mu=2 \ {\rm GeV}) )^{1/3} = 282(10) {\rm MeV}.
\label{eq:sigma13}
\ee

All of these analyses ignore $\rho$. If we knew $f_\pi$, we could divide it out. This
 would reduce
$\Sigma$ by a factor 1/1.42 if $f_\pi=93$ MeV. Since we have no lattice determination
of $f_\pi$ from our simulation, $\rho$ represents a systematic excess, which we must quote separately,
31 MeV for $\Sigma^{1/3}$, for example.

%%%%%%%%%%%%%%%%%%%%%%%%%%%%%%%%%%%%%%%%%%%%%%%%%%%%%%%%%%%%%%%%%%%%%%
\section{Conclusions}
%%%%%%%%%%%%%%%%%%%%%%%%%%%%%%%%%%%%%%%%%%%%%%%%%%%%%%%%%%%%%%%%%%%%%%
A summary of previous calculations of the condensate has recently been given
 by McNeile\cite{McNeile:2005pd}.
Comparing the summary of quenched
determinations there, a three-flavor
prediction by McNeile, our recent $N_f=1$ result \cite{DeGrand:2006uy} of 
$( \Sigma(\overline{MS},\mu=2 \ {\rm GeV}) )^{1/3} = 0.269(9)$ GeV 
(which also does not include $\rho$)
and our $N_f=2$ result quoted above,
the condensate seems to be a quantity which is not very $N_f$ dependent.
Our result is also quite consistent with the Gell-Mann-Oakes-Renner relation, the pion mass and (non-lattice)
phenomenological estimates of the up and down quark masses\cite{Jamin:2002ev}.

This project suffers from a too-small simulation volume, relatively low statistics and we just have 
a single  coarse lattice spacing. The small volume is reflected
in the limited range of eigenmodes which are well-predicted by RMT and by the factor $\rho$
which is present as a systematic in our final answer.  (Recall, $\rho$ approaches unity like $1/L^2$).
In principle, one could determine $f_\pi$ by some other method and scale $\rho$ out of the RMT fit.
We also have the usual problem of a lattice simulation at one value of the lattice spacing:
scaling violations are unknown. In hindsight, the three masses we did here
are overkill; a single small quark mass suffices to give $\Sigma$ from RMT.

In contrast, the strength of this calculation is the use of a chiral lattice fermion, which
allows for the preservation of chiral symmetry in the bare action and insures that spontaneous
chiral symmetry breaking (and explicit chiral symmetry breaking induced by a quark mass)
occurs at finite lattice spacing exactly as in the continuum.  From a simulation
point of view, critical slowing down turned out to be largely eliminated due
to the possibility to run in the opposite chirality sector. 
The definition of the topological charge by the index of the Dirac operator
allows us to easily monitor the topological charge. This makes it possible to
produce ensembles at fixed topology which is convenient for the comparison to 
the predictions of RMT, which are at definite topological charge.

%%%%%%%%%%%%%%%%%%%%%%%%%%%%%%%%%%%%%%%%%%%%%%%%%%%%%%%%%%%%%%%%%%%%%%
\section*{Acknowledgments}
%%%%%%%%%%%%%%%%%%%%%%%%%%%%%%%%%%%%%%%%%%%%%%%%%%%%%%%%%%%%%%%%%%%%%%

This work was supported in part by the US Department of Energy. 
We would like to thank Christian Lang for pointing out a mistake in a previous
version of this paper.


\begin{thebibliography}{99}



%\cite{McNeile:2005pd}
\bibitem{McNeile:2005pd}
  C.~McNeile,
  %``An estimate of the chiral condensate from unquenched lattice QCD,''
  Phys.\ Lett.\ B {\bf 619}, 124 (2005)
  [arXiv:hep-lat/0504006].
  %%CITATION = HEP-LAT 0504006;%%



%\cite{Neuberger:1997fp}
\bibitem{Neuberger:1997fp}
H.~Neuberger,
%``Exactly massless quarks on the lattice,''
Phys.\ Lett.\ B {\bf 417}, 141 (1998)
[arXiv:hep-lat/9707022].
%%CITATION = HEP-LAT 9707022;%%

%\cite{Neuberger:1998my}
\bibitem{Neuberger:1998my}
H.~Neuberger,
%``A practical implementation of the overlap-Dirac operator,''
Phys.\ Rev.\ Lett.\  {\bf 81}, 4060 (1998)
[arXiv:hep-lat/9806025].
%%CITATION = HEP-LAT 9806025;%%



%\cite{Ginsparg:1981bj}
\bibitem{Ginsparg:1981bj}
  P.~H.~Ginsparg and K.~G.~Wilson,
  %``A Remnant Of Chiral Symmetry On The Lattice,''
  Phys.\ Rev.\ D {\bf 25}, 2649 (1982).
  %%CITATION = PHRVA,D25,2649;%%

\bibitem{DeGrand:2005vb}
T.~DeGrand and S.~Schaefer,
Phys.\ Rev.\ D {\bf 72}, 054503 (2005)
[arXiv:hep-lat/0506021].
%%CITATION = HEP-LAT 0506021;%%

%\cite{DeGrand:2006uy}
\bibitem{DeGrand:2006uy}
  T.~DeGrand, R.~Hoffmann, S.~Schaefer and Z.~Liu,
  %``Quark condensate in one-flavor QCD,''
  Phys.\ Rev.\ D {\bf 74}, 054501 (2006)
  [arXiv:hep-th/0605147].
  %%CITATION = HEP-TH 0605147;%%



\bibitem{Shuryak:1992pi}
E.~V.~Shuryak and J.~J.~M.~Verbaarschot,
Nucl.\ Phys.\ A {\bf 560}, 306 (1993) [arXiv:hep-th/9212088].
%%CITATION = HEP-TH 9212088;%%

\bibitem{Verbaarschot:1993pm}
J.~J.~M.~Verbaarschot and I.~Zahed,
Phys.\ Rev.\ Lett.\  {\bf 70}, 3852 (1993) [arXiv:hep-th/9303012].
%%CITATION = HEP-TH 9303012;%%

\bibitem{Verbaarschot:1994qf}
J.~J.~M.~Verbaarschot,
Phys.\ Rev.\ Lett.\  {\bf 72}, 2531 (1994) [arXiv:hep-th/9401059].
%%CITATION = HEP-TH 9401059;%%

\bibitem{Berbenni-Bitsch:1997tx}
M.~E.~Berbenni-Bitsch, S.~Meyer, A.~Sch\"afer, J.~J.~M.~Verbaarschot and T.~Wettig,
Phys.\ Rev.\ Lett.\  {\bf 80}, 1146 (1998) [arXiv:hep-lat/9704018].
%%CITATION = HEP-LAT 9704018;%%



\bibitem{Damgaard:1998ie}
P.~H.~Damgaard, U.~M.~Heller and A.~Krasnitz,
Phys.\ Lett.\ B {\bf 445}, 366 (1999) [arXiv:hep-lat/9810060].
%%CITATION = HEP-LAT 9810060;%%

\bibitem{Gockeler:1998jj}
M.~G\"ockeler, H.~Hehl, P.~E.~L.~Rakow, A.~Sch\"afer and T.~Wettig,
Phys.\ Rev.\ D {\bf 59}, 094503 (1999)
[arXiv:hep-lat/9811018].
%%CITATION = HEP-LAT 9811018;%%

\bibitem{Edwards:1999ra}
R.~G.~Edwards, U.~M.~Heller, J.~E.~Kiskis and R.~Narayanan,
Phys.\ Rev.\ Lett.\  {\bf 82}, 4188 (1999)
[arXiv:hep-th/9902117].
%%CITATION = HEP-TH 9902117;%%

\bibitem{Giusti:2003gf}
L.~Giusti, M.~L\"uscher, P.~Weisz and H.~Wittig,
JHEP {\bf 0311}, 023 (2003) [arXiv:hep-lat/0309189].
%%CITATION = HEP-LAT 0309189;%%


%\cite{Follana:2005km}
\bibitem{Follana:2005km}
  E.~Follana, A.~Hart, C.~T.~H.~Davies and Q.~Mason  [HPQCD Collaboration],
  %``The low-lying Dirac spectrum of staggered quarks,''
  Phys.\ Rev.\ D {\bf 72}, 054501 (2005)
  [arXiv:hep-lat/0507011].
  %%CITATION = HEP-LAT 0507011;%%

%\cite{Wennekers:2005wa}
\bibitem{Wennekers:2005wa}
  J.~Wennekers and H.~Wittig,
  %``On the renormalized scalar density in quenched QCD,''
  JHEP {\bf 0509}, 059 (2005)
  [arXiv:hep-lat/0507026].
  %%CITATION = HEP-LAT 0507026;%%

\bibitem{Berbenni-Bitsch:1998sy}
M.~E.~Berbenni-Bitsch, S.~Meyer and T.~Wettig,
Phys.\ Rev.\ D {\bf 58}, 071502 (1998)
[arXiv:hep-lat/9804030].
%%CITATION = HEP-LAT 9804030;%%

\bibitem{Damgaard:2000qt}
P.~H.~Damgaard, U.~M.~Heller, R.~Niclasen and K.~Rummukainen,
Phys.\ Lett.\ B {\bf 495}, 263 (2000) [arXiv:hep-lat/0007041].
%%CITATION = HEP-LAT 0007041;%%


\bibitem{Damgaard:2000ah}
P.~H.~Damgaard and S.~M.~Nishigaki,
Phys.\ Rev.\ D {\bf 63}, 045012 (2001) [arXiv:hep-th/0006111].
%%CITATION = HEP-TH 0006111;%%

\bibitem{Bietenholz:2003mi}
W.~Bietenholz, K.~Jansen and S.~Shcheredin,
JHEP {\bf 0307}, 033 (2003) [arXiv:hep-lat/0306022].
%%CITATION = HEP-LAT 0306022;%%

\bibitem{Gasser:1986vb}
  J.~Gasser and H.~Leutwyler,
  %``Light Quarks At Low Temperatures,''
  Phys.\ Lett.\ B {\bf 184}, 83 (1987);
  %%CITATION = PHLTA,B184,83;%%
  %``Spontaneously Broken Symmetries: Eeffective Lagrangians At Finite Volume,''
  Nucl.\ Phys.\ B {\bf 307}, 763 (1988).
  %%CITATION = NUPHA,B307,763;%%
See also
  P.~Hasenfratz and H.~Leutwyler,
  %``Goldstone Boson Related Finite Size Effects In Field Theory And Critical
  %Phenomena With O(N) Symmetry,''
  Nucl.\ Phys.\ B {\bf 343}, 241 (1990).
  %%CITATION = NUPHA,B343,241;%%



\bibitem{Bode:1999dd}
A.~Bode, U.~M.~Heller, R.~G.~Edwards and R.~Narayanan,
arXiv:hep-lat/9912043.
%%CITATION = HEP-LAT 9912043;%%


%\cite{Cundy:2005mr}
\bibitem{Cundy:2005mr}
  N.~Cundy,
  %``Current status of dynamical overlap project,''
  Nucl.\ Phys.\ Proc.\ Suppl.\  {\bf 153}, 54 (2006)
  [arXiv:hep-lat/0511047].
  %%CITATION = HEP-LAT 0511047;%%

\bibitem{ouralg}
  T.~DeGrand and S.~Schaefer,
  %``Simulating an arbitrary number of flavors of dynamical overlap fermions,''
  JHEP {\bf 0607}, 020 (2006)
  [arXiv:hep-lat/0604015].
  %%CITATION = HEP-LAT 0604015;%%


%\cite{Sommer:1993ce}
\bibitem{Sommer:1993ce}
  R.~Sommer,
  %``A New way to set the energy scale in lattice gauge theories and its
  %applications to the static force and alpha-s in SU(2) Yang-Mills theory,''
  Nucl.\ Phys.\ B {\bf 411}, 839 (1994)
  [arXiv:hep-lat/9310022].
  %%CITATION = HEP-LAT 9310022;%%


%\cite{Martinelli:1994ty}
\bibitem{Martinelli:1994ty}
  G.~Martinelli, C.~Pittori, C.~T.~Sachrajda, M.~Testa and A.~Vladikas,
  %``A General method for nonperturbative renormalization of lattice
  %operators,''
  Nucl.\ Phys.\ B {\bf 445}, 81 (1995)
  [arXiv:hep-lat/9411010].
  %%CITATION = HEP-LAT 9411010;%%
%G.Martinelli, C.Pittori, C.T.Sachrajda, M.Testa and A. Vladikas, Nucl.
%Phys. B445 (1995) 81.


\bibitem{Duane:1987de}
S.~Duane, A.~D.~Kennedy, B.~J.~Pendleton and D.~Roweth,
Phys.\ Lett.\ B {\bf 195}, 216 (1987).
%%CITATION = PHLTA,B195,216;%%


\bibitem{Fodor:2003bh}
Z.~Fodor, S.~D.~Katz and K.~K.~Szabo,
JHEP {\bf 0408}, 003 (2004) [arXiv:hep-lat/0311010].
%%CITATION = HEP-LAT 0311010;%%



\bibitem{DeGrand:2004nq}
T.~DeGrand and S.~Schaefer,
Phys.\ Rev.\ D {\bf 71}, 034507 (2005) [arXiv:hep-lat/0412005].
%%CITATION = HEP-LAT 0412005;%%



%\cite{Schaefer:2005qg}
\bibitem{Schaefer:2005qg}
  S.~Schaefer and T.~A.~DeGrand,
  %``Dynamical overlap fermions: Techniques and results. Simulations and physics
  %results,''
  PoS {\bf LAT2005}, 140 (2005)
  [arXiv:hep-lat/0508025].
  %%CITATION = HEP-LAT 0508025;%%



\bibitem{Luscher:1984xn}
M.~L\"uscher and P.~Weisz,
Commun.\ Math.\ Phys.\  {\bf 97}, 59 (1985)
[Erratum-ibid.\  {\bf 98}, 433 (1985)].
%%CITATION = CMPHA,97,59;%%

%\cite{Alford:1995hw}
\bibitem{Alford:1995hw}
M.~G.~Alford, W.~Dimm, G.~P.~Lepage, G.~Hockney and P.~B.~Mackenzie,
Phys.\ Lett.\ B {\bf 361}, 87 (1995)
[arXiv:hep-lat/9507010].
%%CITATION = HEP-LAT 9507010;%%

\bibitem{Morningstar:2003gk}
C.~Morningstar and M.~J.~Peardon,
Phys.\ Rev.\ D {\bf 69}, 054501 (2004)
[arXiv:hep-lat/0311018].
%%CITATION = HEP-LAT 0311018;%%

\bibitem{Hasenbusch:2001ne}
M.~Hasenbusch,
Phys.\ Lett.\ B {\bf 519}, 177 (2001)
[arXiv:hep-lat/0107019].
%%CITATION = HEP-LAT 0107019;%%

\bibitem{NR}
Numerical Recipes in C:  The art of scientific computing,   2nd ed,
W.H.~Press, S.A.~Teukolsky, W.T.~Vetterling and B.P.~Flannery,   Cambridge Univ. Pr., 1995.  




%\cite{DeGrand:2005af}
\bibitem{DeGrand:2005af}
  T.~DeGrand and Z.~F.~Liu,
  %``Renormalization of bilinear quark operators for overlap fermions,''
  Phys.\ Rev.\ D {\bf 72}, 054508 (2005)
  [arXiv:hep-lat/0507017].
  %%CITATION = HEP-LAT 0507017;%%

%\cite{Franco:1998bm}
\bibitem{Franco:1998bm}
  E.~Franco and V.~Lubicz,
  %``Quark mass renormalization in the MS-bar and RI schemes up to the NNLO
  %order,''
  Nucl.\ Phys.\ B {\bf 531}, 641 (1998)
  [arXiv:hep-ph/9803491].
  %%CITATION = HEP-PH 9803491;%%

%\cite{Chetyrkin:1999pq}
\bibitem{Chetyrkin:1999pq}
  K.~G.~Chetyrkin and A.~Retey,
  %``Renormalization and running of quark mass and field in the regularization
  %invariant and MS-bar schemes at three and four loops,''
  Nucl.\ Phys.\ B {\bf 583}, 3 (2000)
  [arXiv:hep-ph/9910332].
  %%CITATION = HEP-PH 9910332;%%

%\cite{Hasenfratz:2001hp}
\bibitem{Hasenfratz:2001hp}
  A.~Hasenfratz and F.~Knechtli,
  %``Flavor symmetry and the static potential with hypercubic blocking,''
  Phys.\ Rev.\ D {\bf 64}, 034504 (2001)
  [arXiv:hep-lat/0103029].
  %%CITATION = HEP-LAT 0103029;%%



%\cite{Hasenfratz:2001tw}
\bibitem{Hasenfratz:2001tw}
  A.~Hasenfratz, R.~Hoffmann and F.~Knechtli,
  %``The static potential with hypercubic blocking,''
  Nucl.\ Phys.\ Proc.\ Suppl.\  {\bf 106}, 418 (2002)
  [arXiv:hep-lat/0110168].
  %%CITATION = HEP-LAT 0110168;%%

%\cite{Brodsky:1982gc}
\bibitem{Brodsky:1982gc}
  S.~J.~Brodsky, G.~P.~Lepage and P.~B.~Mackenzie,
  %``On The Elimination Of Scale Ambiguities In Perturbative Quantum
  %Chromodynamics,''
  Phys.\ Rev.\ D {\bf 28}, 228 (1983).
  %%CITATION = PHRVA,D28,228;%%


%\cite{Jamin:2002ev}
\bibitem{Jamin:2002ev}
  M.~Jamin,
%   ``Flavour-symmetry breaking of the quark condensate and chiral  corrections
  %to the Gell-Mann-Oakes-Renner relation,''
  Phys.\ Lett.\ B {\bf 538}, 71 (2002)
  [arXiv:hep-ph/0201174].
  %%CITATION = HEP-PH 0201174;%%

\end{thebibliography}
\end{document}